\documentclass[12pt]{article}
\usepackage{amsmath,amssymb,amscd,cite}
\usepackage{soul,color}
\usepackage{hyperref}
\newtheorem{thm}{Theorem}[section]
\newtheorem{prop}[thm]{Proposition}
\newtheorem{lem}[thm]{Lemma}
\newtheorem{cor}[thm]{Corollary}

\newtheorem{defi}[thm]{Definition}
\newcommand{\pf}{{\bf Proof. \ }}
\newcommand{\qed}{\hfill $\Box$ \\}

\font\msbm=msbm10 at 12pt
\newcommand{\Z}{\mbox{\msbm Z}}

\newcommand{\F}{\mbox{\msbm F}}
\newtheorem{rem}[thm]{Remark}
\newtheorem{ex}[thm]{Example}

\date{}
\begin{document}
\title{Construction of Isodual Quasi-cyclic Codes over Finite Fields}
\author{Fatma-Zahra Benahmed, Kenza Guenda,  Aicha Batoul and T. Aaron Gulliver \thanks{F.-Z. Benahmed and A. Batoul are with the Faculty of Mathematics USTHB, University of Science and Technology of Algiers, Algeria.
  K. Guenda and T. A. Gulliver
are with the Department of Electrical and Computer Engineering, University
of Victoria, PO Box 1700, STN CSC, Victoria, BC, Canada V8W 2Y2,
tel: +250-721-6028, email: kguenda@uvic.ca, agullive@ece.uvic.ca, ORCID 0000-0002-1482-7565, 0000-0001-9919-0323.}}
\maketitle

\begin{abstract}
This paper considers the construction of isodual quasi-cyclic codes.
First we prove that two quasi-cyclic codes are permutation equivalent if and only if their constituent codes are equivalent.
This gives conditions on the existence of isodual quasi-cyclic codes.
Then these conditions are used to obtain isodual quasi-cyclic codes.
We also provide a construction for isodual quasi-cyclic codes as the matrix product of isodual codes.
\hfill 
\hfill 
\end{abstract}

\section{Introduction}

An isodual code is a linear code which is equivalent to its dual, and a self-dual code is a code which is equal to its dual.
The class of isodual codes is important in coding theory because it contains the self-dual codes as a subclass.
In addition, isodual codes are contained in the larger class of formally self-dual codes,
and they are important due to their relationship to isodual lattice constructions~\cite{bachoc}.
Motivated by the numerous practical applications of code equivalency in code-based cryptography~\cite{McEliece,otmani,sendrier1},
we prove that two quasi-cyclic codes are permutation equivalent if and only if their constituent codes are equivalent.
This gives conditions on the existence of isodual quasi-cyclic codes.
These conditions are used to obtain isodual quasi-cyclic codes.
Further, we provide a construction of isodual quasi-cyclic codes as the matrix product of isodual codes.

The remainder of this paper is organized as follows.
In Section 2, some definitions and preliminary results are given.
The main result is given in Section 3.
It is proven that two quasi-cyclic codes are permutation equivalent
if and only if their constituent codes are permutation equivalent.
In Section 4, multiplier equivalent cyclic codes are introduced.
Further, the equivalence of quasi-cyclic codes with cyclic constituent codes is examined.
Then, conditions on the existence of isodual quasi-cyclic codes is considered in Section 5.
In Section 6, the previous results are used to construct quasi-cyclic isodual codes. Namely we give construction of isodual quasi-cyclic codes as matrix product codes using the Vandermonde matrix.


\section{Preliminaries}

Let $C$ be a linear code of length $n$ over a finite field
$\mathbb{F}_q$ and $\sigma$ a permutation of the symmetric group
$S_n$ acting on $\{0,1,\ldots, n-1\}$.
We associate with this code a linear code $\sigma(C)$ defined by
\[
\sigma(C)=\{ (x_{\sigma ^{-1}(0)}, \ldots, x_{\sigma ^{-1}(n-1)}),\, (x_0, \ldots x_{n-1})\in C \}.
\]
We say that the codes $C$ and $C'$ are permutation equivalent if
there exists a permutation $\sigma \in S_n$ such that $C'=\sigma(C)$.
The permutation group of $C$ is the subgroup of $S_n$ given by
\[
Per(C)=\{\sigma \in S_n; \, \sigma(C)=C\}.
\]

A linear code $C$ of length $n$ over $\mathbb{F}_q$ is called
quasi-cyclic (QC) of index $l$ (or an $l$-quasi-cyclic code), if its automorphism group
contains the permutation $T^l$ given by
\begin{equation}
\begin{split}
T^l : \Z_n & \longrightarrow  \Z_n \\
          i &\longmapsto i +l \bmod n.
\end{split}
\end{equation}
This definition is equivalent to saying that for all $c\in {C}$ we have $T^l(c) \in {C}$ where $T : i \mapsto i+1$ is the circular shift.
The index $l$ of ${C}$ is the smallest integer satisfying this property.
If $l=1$, $C$ is called a cyclic code.
From the fact that the permutation group of a cyclic code contains the cyclic shift $T$,
we obtain that this code is an ideal of the ring $\F_q[x]/\langle x^n-1 \rangle$.
Hence it is generated by a polynomial $f(x) | (x^n-1)$.
For a primitive element $\alpha $ of $\F_q$, the defining set of a cyclic code is a subset of
$\Z_n$, given by $S=\{i\le n, f(\alpha^i)=0\}$.
There is a one-to-one correspondence between the irreducible factors of $x^n-1$ and subsets of $S$.
These subsets are called cyclotomic classes.

Let $a$ and $n$ be positive integers such that $\gcd(a,n)=1$.
The permutation $\mu_a$ defined on $\Z_n=\{0,1,\ldots,n\}$ by
\begin{equation}
\begin{array}{ccl}
\label{eq:ling}
\mu_a:\Z_n&\longrightarrow &
 \Z_n\\
 i&\mapsto & \mu_a(i) =ia,
 \end{array}
\end{equation}
is called a multiplier.
Two codes $C$ and $C'$ are called multiplier equivalent if $C'=\mu_a(C)$ for some multiplier $\mu_a$.
Multipliers play an essential role in code equivalence~\cite{G-G}.
The multiplier given in (\ref{eq:ling}) is a special type of permutation which characterizes the
equivalence of some codes.
Multipliers also act on polynomials in $\F_q[x]$ and this gives the following ring automorphism
\begin{equation}
\begin{array}{ccl}
\label{eq:ling2}
\mu_a:\F_q[x]/\langle x^n-1\rangle &\longrightarrow &
 \F_q[x]/\langle x^n-1\rangle\\
 f(x)&\mapsto & \mu_a(f(x)) =
f(x^a).
\end{array}
\end{equation}

If $C$ is a cyclic code generated by $f(x)$, then $\mu_a(C)=\langle f(x^a) \rangle$.
Thus, two cyclic codes $C= \langle f(x) \rangle$ and $D = \langle g(x) \rangle$ are multiplier equivalent
if there exists a multiplier $\mu_a$ such that $g(x)=\mu(f(x))=f(x^a)$.
This definition of multiplier equivalence is more general than that given in \cite{sole3}
as the map $\sigma^e$ where $\sigma(x)=x^q$ was used to define multiplier equivalence.

We attach the standard inner product to $\F_{q}^n$
\[
[{v},{w}] = \sum v_iw_i.
\]
The Euclidean dual code $C^\perp$ of $C$ is defined as
\begin{equation}
C^\perp=\{ {v} \in \F_{q}^n, \ [{v},{w}]= 0 \mbox{ for all } w \in C\}.
\end{equation}
If $C \subseteq C^\perp$, the code is said to be self-orthogonal and
if $C=C^\perp$ the code is self-dual.
A linear code which is equivalent to its dual is called an isodual code.

Let $f(x)= a_0+a_1x+\ldots +a_rx^r$ be a polynomial of degree $r$
with $f(0)= a_0\neq 0$.
Then the monic reciprocal polynomial of $f(x)$ is
\[
f^*(x)= f(0)^{-1}x^rf(x^{-1}) = a_0^{-1}(a_r+a_{r-1}x+\ldots +a_0x^r).
\]
A self-reciprocal polynomial is a polynomial that is equal to its reciprocal.
\section{Equivalent Quasi-Cyclic Codes}
In this section, we characterize the problem of permutation equivalence for quasi-cyclic codes over finite fields.
Let $\F_q$ be the finite field of cardinality $q$ and $m$ be a positive integer such that $\gcd(m,q)=1$.
Further, let $\F_{q}[Y]$ denote the ring of polynomials in the indeterminate $Y$ over $\F_q$.
Consider the ring $R=\F_{q}[Y]/\langle Y^m-1\rangle$.
For a positive integer $l$, define the map
\begin{equation}
\label{eq:decom}
\begin{split}
\Phi :\F_{q}^{lm}  &  \longrightarrow R^l  \\
c=(c_{0,0},c_{0,1},\ldots, c_{0,l-1},\ldots,c_{m-1,0}, \ldots, c_{m-1,l-1}) &\longmapsto
 \Phi (c)= (c_0(Y),c_1(Y), \ldots, c_{l-1}(Y)),
\end{split}
\end{equation}
where $c_j(Y)= \sum \limits_{i=0}^{m-1}c_{i,j}Y^i \in \F_{q}$.
It was shown in~\cite{sole1} that the map $\Phi$ induces a one-to-one
correspondence between QC codes over $\F_{q}$ of index $l$ and
length $lm$, and linear codes over $R$ of length $l$ where each codeword coordinate is a polynomial of degree at most $m-1$.
In~(\ref{eq:decom}), each coordinate $c_{i,j}$ in
$c=(c_{0,0},\ldots, c_{0,l-1},\ldots ,c_{m-1,0}, \ldots, c_{m-1,l-1})$
can be written as $c_{j+il}$, $0\le j\le l-1$, $0\le i \le m-1$, so
$c_j(Y)=\sum \limits_{i=0}^{m-1}c_{ij}Y^i \in R$ can be expressed in vector form as
$c_j(Y)=(c_{0,j}, c_{1,j},\ldots, c_{m-1,j})$.
Then the image of the codeword $(c_{j+il})_{ 0\le j\le l-1, 0\le i \le m-1}$
by the map $\Phi$ is the codeword $(c_{i+jm})_{ 0\le j\le l-1, 0\le i \le m-1}$.
This implies the following result.
\begin{prop}
\label{prop:image}
Let $\mathcal{C}$ and $\mathcal{C}'$ be quasi-cyclic codes of length $n=lm$ and index $l$ over $\F_q$.
Then $\mathcal{C}$ and $\mathcal{C}'$ are permutation equivalent if and only if the codes
$C=\Phi(\mathcal{C})$ and $C'=\Phi(\mathcal{C})$ are permutation equivalent.
\end{prop}
\pf
Assume that $\mathcal{C}=\{(c_{j+il})_{ 0\le j\le l-1, 0\le i \le
m-1}\}$ and $\mathcal{C}'=\{(c'_{j+il})_{ 0\le j\le l-1, 0\le i \le m-1}\}$
are equivalent by a permutation $\sigma \in S_n$.
Hence if $\sigma$ is such that $\sigma (j+il)= j'+i'l$ we have
\[
\sigma((c_{j+il})_{0\le j\le l-1,0\le i \le m-1}) = (c'_{j+il})_{ 0\le j\le l-1,0\le i \le m-1} = (c_{j'+i'l})_{ 0\le j'\le l-1,0\le i' \le m-1},
\]
and therefore
\[
\Phi(\sigma((c_{j+il})_{0\le j\le l-1,0\le i \le m-1})) =\Phi ((c_{j'+i'l})_{ 0\le j'\le l-1,0\le i' \le m-1})= (c_{i'+j'm})_{ 0\le j'\le l-1,0\le i' \le m-1},
\]
with the associated permutation $\tau$ given by $\tau(i'+j'm)=i+jm$.
Since $\sigma$ is in $S_n$, $\tau$ is also in $S_n$.
In addition, $\tau$ is such that $\tau(\Phi(\sigma(\mathcal{C}))=\Phi(\mathcal{C})$.
This proves the first implication.

Now assume that $C=\{(c_{i+jm})_{ 0\le j\le l-1, 0\le i \le m-1}\}$ and $C'=\{(c'_{i+jm})_{ 0\le j\le l-1, 0\le i \le m-1}\}$ are
images by the map $\Phi$ of two QC codes $\mathcal{C}$ and $\mathcal{C'}$, respectively,
and there exists a permutation $\sigma$ such that
\[
\sigma (C)=\sigma(\{(c_{i+jm})_{ 0\le j\le l-1, 0\le i \le m-1} \})=C'=\{(c'_{i+jm})_{ 0\le j\le l-1, 0\le i \le m-1}
\}=\{(c_{i'+j'm})_{0\le j\le l-1, 0\le i \le m-1}\}.
\]
Hence
$\mathcal{C}= \{(c_{j+il})_{0\le j\le l-1, 0\le i \le m-1}\}$ and
$\mathcal{C'}= \{(c_{j'+i'l})_{0\le j'\le l-1, 0\le i' \le m-1}\}$.
Then by defining the permutation
$\tau$ such that $\tau(j'+i'l)=j+il$ we obtain
$\tau(\mathcal{C}')=\mathcal{C}$.
\qed

Now we consider the factorization of $Y^m-1$ over $\F_q$.
Since it is assumed that $\gcd (m,q)=1$, then
$Y^m-1$ has a unique decomposition into irreducible factors over $\F_q$
\begin{equation}
\label{eq:fact1} Y^m-1=\delta g_1\ldots g_s h_1h_1^{*} \ldots h_t h_t^*,
\end{equation}
where $\delta$ is a unit in $\F_{q}$, $h_i^{*}$ is the reciprocal of $h_i$, and $g_i$ is self-reciprocal.
The ring $R$ is a principal ideal ring, so it can be decomposed into a direct sum of local rings.
Hence the Chinese Remainder Theorem gives the following decomposition
\begin{equation}
\label{eq:decom2}
R=\frac{\F_{q}[Y]}{\langle Y^m-1 \rangle}=\left(\bigoplus_{i=1}^s
\frac{\F_{q}[Y]}{\langle g_i\rangle}\right)\bigoplus\left(\bigoplus_{j=1}^t\left(\frac{\F_{q}[Y]}{\langle h_j\rangle}
\bigoplus
\frac{\F_{q}[Y]}{ \langle h_j^*\rangle}\right)\right).
\end{equation}
Let $G_i=\frac{\F_{q}[Y]}{\langle g_i\rangle}$, $H_j'=\frac{\F_{q}[Y]}{\langle h_j\rangle}$ and $H_j''=\frac{\F_{q}[Y]}{\langle{h_j}^*\rangle}$.
Since the polynomials in the decomposition (\ref{eq:fact1}) are irreducible, then
the local rings are in fact field extensions of $\F_q$.
Hence as a consequence of the decomposition (\ref{eq:decom2}), we obtain
that every $R$-linear code of length $l$ can be decomposed as
$C=(\oplus_{i=1}^s C_i )\oplus(\oplus_{j=1}^t (C_j'\oplus C_j''))$,
where $C_i$ is a linear code over $G_i$,
$C_j'$ is a linear code over $H_j'$,
and $C_j''$ is a linear code over $H_j''$.
The codes $C_i$, $C_j$ and $C_j''$
are called the {\it components} of the QC code $\mathcal{C}$.

Assume that $g_i$ is one of the self-reciprocal polynomials in~(\ref{eq:fact1}).
We now consider the action of the following map over the local component ring $\F_{q}[Y]/\langle g_i\rangle=G_i$ of $R$
\begin{equation}
\begin{array}{ccl}
\label{eq:ling3} \,-\,  :\F_{q}[Y]/\langle g_i\rangle
&\longrightarrow &
\F_{q}[Y]/\langle g_i\rangle\\
c(Y)&\mapsto & c(Y^{-1}).
\end{array}
\end{equation}
The map $-$ is a ring automorphism.
For $g_i$ of degree 1 this map is the identity and if $\deg(g_i)=K_i \neq 1$, we know that $g_i$ and $g_i^*$ are associates, so $K_i$ must be even.
Since $g_i$ is irreducible and square free, it is also separable and a local polynomial
($g_i$ is called a local polynomial  if $\F_{q}[Y]/\langle g_i\rangle$ is a local ring \cite{Mac2}).
Further, as $g_i$ is irreducible of degree $K_i$,
from \cite[Theorem 4.2]{Mac2} the ring $G_i=\F_{q}[Y]/\langle g_i \rangle$ is an extension of $\F_q$,
namely $\F_{q^{K_i}}$.
Then the map $r\mapsto \overline{r}$, is the map $\nu :r \mapsto r^{q^{K_i/2}}$, and hence is a power of the Frobenius map.
Thus, it is a permutation of $\F_{q^{K_i}}$ which fixes the elements of $\F_{q}$.
This proves the following result.
\begin{lem}
\label{lem:eqH}
With the notation above, each code $C_i$ over $G_i$ is equivalent to $\nu(C_i)=\overline{C_i}$.
\end{lem}

For $a=(a_0,\ldots,a_{l-1})$ and $b=(b_0,\ldots,b_{l-1})$ in $G_{i}^l$,
the Hermitian inner product on $G_i$ is defined as
\begin{equation}
\label{eq:Herm}
\langle a,b\rangle^{H}= \sum\limits^{l-1}_{k=0}a_{k}\overline{b_{k}},
\end{equation}
This is in fact the usual Hermitian inner product.

\begin{lem}
\label{lem:mono}
Let $C_i$ be a linear code over $G_i$.
The Hermitian dual of $C_i$ denoted ${C_i}^{\bot H}$ is equivalent to the Euclidean dual of $C_i$.
\end{lem}
\pf
Define the code $\overline{C}=\{\overline{r}, \, r\in C\}$. It is
easy to show that $ {C_i}^{\bot H}=\overline{(C_i)}^{\bot}=\nu (C_i)^{\bot}$.
Hence from Lemma~\ref{lem:eqH} we have that $\nu (C_i)^{\bot} = (\nu (C_i^{\bot})$.
\qed

For $a = (a_{1},a_{2}...a_{lm}) , b = (b_{1},b_{2}...b_{lm}) \in \F_q^{lm}$, let $\Phi(a)=(c_0,\ldots, c_{l-1})$ and $\Phi(b)=(d_0,\ldots, d_{l-1})$,
where
\[
c_i= (c_{i,1}, \ldots, c_{i,s},c'_{i,1}, c''_{i,1}, \ldots, c'_{i,t}c''_{i,t}),
\]
and
\[
d_i= (d_{i,1}, \ldots, d_{i,s},d'_{i,1}, d''_{i,1}, \ldots, d'_{i,t}d''_{i,t}),
\]
with $c_{i,j}, d_{i,j}\in G_j$, $c'_{i,j}, d'_{i,j}\in {H_j}'$, and $c''_{i,j}, d''_{i,j}\in {H_j}''$.

We define the Hermitian inner product on $R^l$ by
\begin{eqnarray}
\langle \Phi(a),\Phi(b) \rangle&=&\left( \sum_i {c_{i,1}} \overline{d_{i,1 }}, \ldots, \sum_i {c_{i,s}} \overline{d_{i,s }}, \right.
\nonumber \\
&&\sum_i c'_{i,1} d''_{i,1 }, \sum_i c''_{i,1}d'_{i,1 }, \ldots
\nonumber \\
&&\left. \sum_i c'_{i,t}d''_{i,t},\sum_i c''_{i,t} d'_{i,t }\right).\nonumber
\end{eqnarray}
Using this inner product, Lim~\cite{lim} and Ling and Sol\'e~\cite{sole1} gave the Euclidean dual of a QC code.
\begin{prop}
\label{propodual}
Let $\mathcal{C}$ be an $l$-QC code of length $n=lm$ over $\F_q$
and $C=\Phi(\mathcal{C})=(\oplus_{i=1}^s C_{i}\oplus(\oplus_{j=1}^t (C_{j}^{'}\oplus C_{j}^{''})))$ be its image as defined previously.
Then the Euclidean dual of $\mathcal{C}$ is the $l$-QC code $\mathcal{C}^{\bot}$ such that
$\Phi(\mathcal{C}^{\bot})=(\oplus_{i=1}^s C_{i}^{\bot H}\oplus(\oplus_{j=1}^t (C_{j}^{''\bot}\oplus C_{j}^{'{\bot}})))$.
\end{prop}

We require the following lemma concerning the direct sum of codes over a commutative ring.
\begin{lem}
\label{lem:direct}
Assume that $C= C_1\oplus C_2$ and $C'= C_1'\oplus C_2'$ are codes of length $2n$ which are the direct
sums of codes of length $n$.
Then there exists a permutation $\sigma \in S_{2n}$ such that $\sigma(C) =C'$ if and only if there exist
permutations $\sigma_1$ and $\sigma_2$ in $S_n$ such that $\sigma_1(C_1)=C_1'$ and $\sigma_2(C_2)=C_2'$.
\end{lem}
\pf
Assume that
\[
\sigma(C)= \sigma(C_1\oplus C_2)= {C}',
\]
and
\[
C'= C_1'\oplus C_2'=\{(c_{\sigma(1)},\ldots c_{\sigma(n)},c_{\sigma(n+1)},\ldots, c_{\sigma(2n)})\}
\]
with
$(c_1,\ldots,c_n)\in C_1$ and $(c_{n+1},\ldots,c_{2n}) \in C_2$.
This gives that $\sigma(i) \in\{1,\ldots,n\}$ for $1\le i\le n$, and
$\sigma(i) \in\{n+1,\ldots,2n\}$ for $n+1\le i\le 2n$.
Hence we can define the permutations $\sigma_1$ and $\sigma_2$ on $n$ elements by
$\sigma_1(1)=\sigma (1), \ldots, \sigma_1(n)= \sigma(n)$, and
$\sigma_2(1)=\sigma(n+1), \ldots, \sigma_2(n)=\sigma(2n)$.
Then $\sigma(C_1\oplus C_2)=\sigma_1(C_1)\oplus \sigma_2(C_2)= C_1'\oplus C_2'$.
Let the mapping $Pr_1$ be the projection on the first $n$ coordinates so that
$Pr_1( \sigma_1(C_1)\oplus \sigma_2(C_2))= \sigma_1(C_1)= Pr_1(C_1'\oplus C_2')=C_1'$ and
then $\sigma_1(C_1)=C_1'$.
We also obtain $\sigma_1(C_2)=C_2'$ by considering the projection $Pr_2$ on the last $n$ coordinates.
For the converse, assume that there exist permutations $\sigma_1$ and $\sigma_2$ such that
$\sigma_1(C_1)=C_1'$ and $\sigma_2(C_2)= C_2'$.
Hence we obtain the permutation $\sigma \in S_{2n}$ given by $\sigma(i)=\sigma_1(i)$
and $\sigma(i+n)=\sigma_2(i)$ for $1 \le i \le n $, so then $\sigma(C)=C'$.
\qed
\begin{rem}
Lemma~\ref{lem:direct} can be easily generalized for the direct sum of $k>2$ codes of the same length.
Further, the result is independent of the structure of the underlying finite commutative ring.
This is due to the action of the permutation on the codes.
\end{rem}
We now give the main result for this section.
\begin{thm}
\label{Thm:isodual}
Let $\mathcal{C}$ be a quasi-cyclic code of length $n=lm$ and index $l$ over $\F_q$ such that
$\Phi(\mathcal{C})=(\oplus_{i=1}^s C_i)\oplus(\oplus_{j=1}^t (C_j'\oplus C_j''))$.
Then $\mathcal{C}$ is isodual if and only if each of its components $C_i$ for $1\le i\le s$ is
isodual and for each $1 \le j \le t$ we have that $C_j'$ is equivalent to $C_j''^{\bot}$.
\end{thm}
\pf
Let $\mathcal{C}$ be an $l$-QC code which is isodual.
Then there exists a permutation $\sigma$ such that $\mathcal{C}=\sigma (\mathcal{C}^{\bot})$.
By Proposition~\ref{prop:image}, there exists a permutation $\tau$ such that
$\Phi(\mathcal{C})=\tau(\Phi(\mathcal{C}^{\bot}))$.
From~Proposition~\ref{propodual} we have that
$\Phi(\mathcal{C}^{\bot}) =\Phi(\mathcal{C})^{\bot H}=(\oplus_{i=1}^s( C_i^{\bot H})\oplus(\oplus_{j=1}^t
(C_j''^{\bot}\oplus C_j'^{\bot})))$.
Hence from Lemma~\ref{lem:direct} there exist permutations $\tau_i$, $\tau_j'$ and $\tau_j''$ such
that $C_i= \tau_i(C_i^{\perp H})$, $C_j'= \tau_j'(C_j'^\perp)$ and
$C_j'=\tau_j''(C_j''^\perp)$.
From Lemma~\ref{lem:mono} we have that ${C_i}^{\bot H}= \nu (C_i)^{\bot}$,
so $C_i= \tau_i(\nu(C_i^{\perp }))$.
Then for $1\le i\le s$, the component $C_i$ is isodual.

For the converse, assume that each component of $\mathcal{C}$ is isodual.
Then we have that $\tau_i(C_i)=C_i^{\bot}$ for $1\le i\le s$,
$\tau_j'( C_j')= C_j'^\bot$ and $\tau_j''(C_j'')=( C_j'')^{\bot}$ for $1\le j\le t$.
From Lemma~\ref{lem:mono} we have that $C_i^{\bot H}=\nu({C_i}^{\bot})$.
Hence $C_i^{\bot H}=\nu(\tau_i{C_i})$ and so
$\Phi(C)^{\bot}=(\oplus_{i=1}^s \nu(\tau_i( C_i))\oplus(\oplus_{i=1}^s \tau_j'( C_j')\oplus \tau_j''(C_j'')))$.
Then from Lemma~\ref{lem:direct} there exists a permutation $\theta \in S_n$ such that
$\Phi(C)^{\bot}= \theta ((\oplus_{i=1}^s C_i)\oplus(\oplus_{j=1}^t (C_j'\oplus C_j''))$, and by
Proposition~\ref{prop:image} $\mathcal{C}$ is isodual.
\qed
\begin{ex}
If $q\equiv 1 \bmod 3$, then $Y^3-1$ can be factored as $(Y-1)(Y-\beta)(Y-\beta^2)$, where $\beta^2+\beta+1=0$ and $\beta\in\F_q$.
Hence, an $l$-QC code $\mathcal{C}$ over $\F_q$ of length $3l$ decomposes into $C_1\oplus C_2 \oplus C_3$,
where $C_1$, $C_2$ and $C_3$ are codes over $\F_q$ of length $l$.
Then by Theorem \ref{Thm:isodual}, $\mathcal{C}$ is isodual if and only if $C_1$ is isodual and
$C_3$ is equivalent to $C_2^{\perp}$ with respect to the Euclidean inner product.
As a special case, for isodual codes $C_1$ and $ C_2$ of length $l$
the code $C_1\oplus C_2 \oplus C_2^{\bot}$ is an isodual $l$-QC code of length $3l$ over $\F_q$.
\end{ex}

The following corollary is a direct consequence of Proposition~\ref{prop:image} and Theorem~\ref{Thm:isodual}.
Note that this result was given in \cite[Theorem 4.2]{sole1}.
\begin{cor}
\label{cor:condi}
An $l$-QC code $\mathcal{C}$ of length $lm$ over $R$ is self-dual if and only if
\[
\Phi(\mathcal{C})=\left(\bigoplus_{i=1}^s C_i\right)\bigoplus\left( \bigoplus_{j=1}^t \left(C_j'\bigoplus (C_j')^{\bot}\right)\right),
\]
where for $1\le i \le s$, $C_i$ is a self-dual code over $\frac{R[Y]}{\langle g_i\rangle}$ with respect to the Hermitian inner product,
and for $1 \le j \le t$, $C_j'$ is a linear code of length $l$ over $H_j$ and $ C_j'^{\bot}$ is its dual with
respect to the Euclidean inner product.
\end{cor}

From Corollary \ref{cor:condi}, it can easily be determined that the index of a self-dual $l$-QC code must be even.
In~\cite[Proposition 6.1]{sole1},
conditions were given on the existence of self-dual QC codes of index $2$.
We now generalize these results to give conditions on the existence of self-dual QC codes of index $l$ even.
\begin{thm}
Let $m$ be an integer relatively prime to $q$.
Then self-dual QC codes over $\F_q$ of length $lm$, with $l$ even exist if
and only if $ (-1)^{l/2}$ is a square in $\F_q$.
\end{thm}
\pf
If a self-dual QC code $\mathcal{C}$ over $\F_q$ of length $lm$ exists, then Corollary~\ref{cor:condi}
shows that there is a self-dual code $C_{i_0}$ of length $l$ over $G_{i_o}$ for $1\leq i_0\leq s$. Hence
by~\cite[Theorem 9.1.3]{huffman03} $(-1)^{l/2}$ is a square in $\F_q$.
Thus, the condition in the theorem is necessary.
Conversely, if $(-1)^{l/2}$ is a square in $\F_q$ then
it is also a square in $\F_{q^{K_{i}}} = G_{i}$ which is an extension field of $\F_q$,
so a self-dual code $C_{i}$ over $G_{i}$ exists 
for all $1\leq i\leq s$.
Then
\[
\left(\bigoplus_{i=1}^s C_i\right)\bigoplus\left( \bigoplus_{j=1}^t \left(C_j'\bigoplus (C_j')^{\bot}\right)\right),
\]
is a self-dual QC code of length $lm$ over $\F_q$ with $C_j'$ a trivial code over $H'_{j}$.
\qed
\begin{ex}
\label{ex:Herm}
For $q=9$ and $m=2$, the polynomial $y^2-1$ factors into distinct linear factors $(y-1)$ and $(y+1)$,
each of which is self-reciprocal.
Hence, $R$ decomposes into the direct sum $\F_q\oplus \F_q$, and an $l$-QC code of length $2l$
over $\F_q$ can be expressed as $C_1\oplus C_2$, where $C_1$ and $C_2$ are codes over $\F_q$ of length $l$.
Then from Corollary~\ref{cor:condi}, if $C_1$ and $C_2$ are Hermitian self-dual codes of length $l$,
then $C_1\oplus C_2$ is an $l$-QC code of length $2l$ which is self-dual.
\end{ex}

\section{Multiplier Equivalent Quasi-Cyclic Codes}
A natural question that arises is, can a multiplier be a permutation by which two quasi-cyclic codes are equivalent?
From Lemma~\ref{lem:direct} and Proposition~\ref{propodual},
we have that two quasi-cyclic codes are equivalent if and only if their constituent codes are equivalent.
Hence we give the following definition.
\begin{defi}
\label{defi:multiplier}
Two quasi-cyclic codes $\mathcal{C}$ and $\mathcal{D}$ are multiplier equivalent if and only if all their components are multiplier equivalent.
\end{defi}
In the next section, conditions are given on when two quasi-cyclic codes with cyclic components are multiplier equivalent.
Further, it will be shown that Definition \ref{defi:multiplier} is more general than that given in \cite{sole3}.
\subsection{Equivalence of Quasi-Cyclic Codes with Cyclic Constituent Codes}
In this section, we consider the equivalence of quasi-cyclic codes with cyclic constituent codes as described in \cite{lim,sole1,solering,sole3},
so $\Phi(\mathcal{C})$ is a cyclic code, i.e. an ideal of $R[X]/\langle X^l-1\rangle$.
We have the following results.
\begin{prop}(\cite[Proposition 8]{lim})
\label{prop:lim}
Let $q$ be a prime power and $\F_q$ the finite field with $q$ elements.
Further, let $l$ and $m$ be positive integers with $m$ coprime to
$q$, and $\mathcal{C}$ be a quasi-cyclic code of length $n=lm$ and index $l$
over $\F_q$.
Then the following are equivalent.
\begin{enumerate}
\item[(i)] $\Phi(\mathcal{C})$ is cyclic.
\item[(ii)] all of the constituent codes of $\mathcal{C}$ are cyclic.
\end{enumerate}
\end{prop}
\begin{thm}
With the assumptions of Proposition \ref{prop:lim}, since $\gcd (m,q)=1$, then
$Y^m-1$ has a unique decomposition into irreducible factors over $\F_q$
\begin{equation}
\label{eq:fact}
 Y^m-1=\delta g_1\ldots g_s h_1h_1^{*} \ldots h_t h_t^*.
\end{equation}
Then the number of quasi-cyclic codes of length $lm$ with cyclic constituent codes is
\[
\prod_{i=1}^{s} 2^{C_{q^{n_i}}(l)}\left(\prod_{j=1}^{t} 2^{C_{q^{n_j}}(l)}\right)^2,
\]
where $C_{q^{n_k}}(l)$ is the number of $q^{n_k}$-cyclotomic classes of $\F_{q^{n_k}}$ modulo $l$,
and $n_k$ is the degree of the irreducible factor $f_k$ over $\F_q$ in the factorization (\ref{eq:fact}).
\end{thm}
\pf
From the decomposition (\ref{eq:decom2}) and Proposition \ref{prop:lim}, it can be deduced that enumerating these quasi-cyclic codes
requires the number of cyclic codes over $G_i$, $H_j'$ and $H_j''$.
Since the polynomials in (\ref{eq:fact}) are irreducible,
the rings $G_i$, $H_j'$ and $H_j''$ are field extensions of $\F_q$ of degree equal to the degree of the corresponding polynomial in (\ref{eq:fact}).
Hence, the problem is reduced to enumerating cyclic codes of length $l$ over $\F_{q^{n_k}}$.
It is well known \cite{huffman03} that this quantity is equal to $2^{C_{q^{n_k}}(l)}$. Then
the result follows.
\qed

\begin{thm}
Let $\mathcal{C}$ and $\mathcal{D}$ be quasi-cyclic codes of length $lm$ and index $l$,
both with cyclic constituent codes
and such that $gcd(l, \phi (l))=1$, where $\phi (\cdot)$ is Euler's phi function.
Then $\mathcal{C}$ and $\mathcal{D}$ are equivalent if and only if they are multiplier equivalent.
\end{thm}
\pf
Assume that $\mathcal{C}$ and $\mathcal{D}$ are quasi-cyclic codes with cyclic constituent codes.
Then from Proposition~\ref{prop:lim} all the constituent codes are cyclic.
Furthermore, from Theorem~\ref{Thm:isodual} $\mathcal{C}$ and $\mathcal{D}$ are equivalent if and only if their
cyclic constituent codes are equivalent.
These cyclic codes have length $l$ over an extension field such that $\gcd(l, \phi (l))=1$.
Then from~\cite[Theorem 1]{job}, $\mathcal{C}$ and $\mathcal{D}$ are equivalent if and only if they are multiplier equivalent,
and the result follows.
\qed
\begin{ex}
If two quasi-cyclic codes with index a prime $p$ and cyclic constituent codes are equivalent,
then they are equivalent only by a multiplier.
\end{ex}

Given a quasi-cyclic code $\mathcal{C}$ with cyclic constituent codes, we now consider the number of quasi-cyclic codes which are equivalent to $\mathcal{C}$.
\begin{thm}
\label{th:prime}
Let $\mathcal{C}$ be a quasi-cyclic code of length $lm$ and index $l$ with cyclic constituent codes such that $gcd(l, \phi (l))=1$.
Then the number of quasi-cyclic codes equivalent to $\mathcal{C}$ is $(\phi (l)+1)^ {\alpha}$
where $\alpha$ is the number of irreducible factors of $Y^m-1$.
\end{thm}
\pf
Under the hypotheses of the theorem, the components $C_i$, $C_j'$ and $C_j''$ of $\mathcal{C}$ are cyclic.
If $\mu_a$ is a multiplier, then the quasi-cyclic code with components $\mu(C_1)$, $C_i$, $i\ne 1$, $C_j'$ and $C_j''$,
is equivalent to $\mathcal{C}$.
This also holds for quasi-cyclic codes with components $C_1$, $\mu_a(C_2)$, $C_i$, $i \ne 2$, $C_j'$ and $C_j''$.
Further, it is true for quasi-cyclic codes with constituent codes $\mu_a(C_{k})$, $k\in \{1,\le s\}$
or $k \in \{1 \le t \}$ and all others equal to $C_i$, $C_j'$ or $C_j''$.
Since there are $\phi (l)$ multipliers and $\alpha$ components,
the number of quasi-cyclic codes equivalent to $\mathcal{C}$ which differ in only one component
$\mu_a(C_k)$ is $\alpha(\phi (l))$, where $\alpha$ is the number of components of $\mathcal{C}$ which is also
the number of factors of $Y^m-1$.
Similarly, the number of equivalent quasi-cyclic codes which differ from $\mathcal{C}$
in only two components $\mu_a(C_k)$ and $\mu_b(C_{h})$ is equal to $\binom{\alpha}{2}(\phi (l))^2$.
Then the number of quasi-cyclic codes equivalent to $\mathcal{C}$ is
\[
\sum_{k=0}^\alpha \binom{\alpha}{k}(\phi (l))^k=(\phi(l)+1)^\alpha.
\]
\qed

\section{Isodual Quasi-Cyclic Codes}
In this section, conditions are given on the existence of isodual quasi-cyclic codes over $\F_q$.
The results are based on the existence of isodual cyclic codes.
Thus, we first consider the existence of these codes.
\begin{prop}\cite[Theorem 4.1]{BBGA}
\label{prop:equivalent}
Let $C$ be a cyclic code of length $n$ over $\F_{q}$ generated by the polynomial $g(x)$ and $\lambda \in \F_{q}^*$
such that $\lambda^n=1$.
Then the following holds:
\begin{enumerate}
\item[(i)] $C$ is equivalent to the cyclic code generated by $g^*(x)$, and
\item[(ii)] $C$ is equivalent to the cyclic code generated by $g(\lambda x)$.
\end{enumerate}
\end{prop}

From Proposition \ref{prop:equivalent}, there are several constructions of isodual cyclic codes over finite fields.
Before providing some applications, we next give conditions on the existence of isodual quasi-cyclic codes.
\begin{thm}
If there exists an isodual quasi-cyclic code of index $l$, then $l$ must be even.
There exist no self-dual or isodual multiplier equivalent quasi-cyclic codes with cyclic constituent codes over $\F_q$ if $q$ is odd.
When $l=2$, there always exists a quasi-cyclic code with cyclic constituent codes which is isodual.
Further, there exists an isodual quasi-cyclic code over $\F_q$ of index $l=2s$ for $s$ odd.
\end{thm}
\pf
From Theorem \ref{Thm:isodual}, a condition on the existence of an isodual quasi-cyclic code
is that the constituent codes $C_i$, $1\le i\le s$, are linear isodual codes of length $l$.
This is possible if and only if $l$ is even.
Now assume there exists a quasi-cyclic code with cyclic constituent codes which is also self-dual or multiplier isodual.
Then from Theorem~\ref{Thm:isodual} and Proposition~\ref{prop:lim}, for $1 \le i \le s$, constituent code $C_i$ must be cyclic and self-dual or multiplier isodual.
It is well known that no cyclic self-dual or multiplier isodual codes exist if $q$ is odd~\cite{jia2}.
Hence there are no QC multiplier equivalent self-dual or multiplier isodual codes in this case.

If $l=2$, then $x^2-1=(x-1)(x+1)$, so from Proposition~\ref{prop:equivalent}(i) the code generated by
$(x-1)$ is equivalent to the code generated by $x+1$, which is its dual.
Consider the quasi-cyclic code with cyclic constituent codes
$C_i =\langle (x-1)f(x) \rangle$ and $C_j'=C_j'' = \langle (x-1)f(x) \rangle$.
Since $C_j'=C_j''$ and they are over the same field extension (the degree of $g$ is the same as that of $g^*$),
the result follows from Theorem~\ref{Thm:isodual}.
\qed
\section{Applications}
We now provide applications of the results given in the previous sections.
The first employs the Vandermonde construction of isodual codes,
and the second is the cubic construction of isodual and self-dual codes.

\subsection {The Vandermonde Construction of Isodual Quasi-Cyclic Codes}
In this section, isodual quasi-cyclic codes are obtained using the Vandermonde construction.
Let $\F_q$ be a finite field with odd characteristic and $m'$ an odd integer such that $\gcd(m',q)=1$.
For an integer $a\geq1$, suppose there exists a $2^a$-th primitive root of unity $\alpha$ in $\F_q$.
Then by \cite[Lemma 3.2]{BGGI}, $\alpha^i$ is a unit for all $1\leq i\leq2^a-1$.
We have the following result on the inverse of a Vandermonde matrix.
Let $V=(\alpha^{ij})_{0\leq i,j\leq2^a-1}$ where $2^a$ and $\alpha$ are as described above.
This matrix is invertible and the inverse has the simple form given in the following lemma.
\begin{lem}
\label{lem:Van}
Let $V=(\alpha^{ij})_{0\leq i,j\leq2^a-1}$ be the Vandermonde matrix described above, then its inverse is
 \[V^{-1}=(2^{a})^{-1}(\alpha^{-ij})_{0\leq i,j\leq2^a-1}.\]
\end{lem}
\pf
For $0\leq i,j\leq 2^a-1$, the $ij$-th entry of $2^aV^{-1}$ is $\sum_{k=0}^{2^a-1}\alpha^{k(i-j)}$.
When $i=j$, this sum is $2^a$, and if $i\neq j$, it is a geometric series with value $(\alpha^{2^a(i-j)}-1)/(\alpha^{i-j}-1)$
which is zero since $\alpha^{2^a}=1$.
\qed

Now we describe the Vandermonde product of codes.
For this, let the vectors $c_0,c_1,\ldots,c_{2^a-1}\in \F_q^{2^am'}$
form the rows of a $2^a\times2^am'$ matrix $Q$.
Further, let $V^{-1}$ be the matrix given in Lemma \ref{lem:Van}.
Define $P=V^{-1}Q$ and denote the rows of $P$ by $u_0,u_1,\ldots,u_{2^a-1}\in \F_q^{2^a}$.
Finally, denote the concatenation of these rows by $u=(u_0|u_1|\ldots|u_{2^a-1})\in \F_q^{2^{2a}m'}$.
For $k=0,1,\ldots,2^a-1$, let $C_k$ be a code of length $2^am'$ over $\F_q$
and define
 \begin{equation}
\label{eq:Vander}
C_0\vee C_1\vee \ldots \vee C_{2^a-1}=\{u:P=V^{-1}Q, \,c_{k}\in C_{k},\,\,\ 0\leq k\leq 2^a-1\}.
\end{equation}
We call this code the Vandermonde product of $C_0, C_1, \ldots, C_{2^a-1}$.
The factor $2^{-a}$ can be ignored because it multiplies all codewords by a unit
and so does not change the code properties.
If $2^{-a}$ is ignored, then $V^{-1}$ is just a permutation of $V$ (swapping rows $i$ and $2^a-i$).

In the following,
we give the connection between the Vandermonde construction of some quasi-cyclic codes and the construction of matrix product codes.
First, we require some results on matrix product codes.
Matrix product codes over finite fields were introduced in \cite{Blackmore}.
This construction includes the Plotkin and Turyn constructions as special cases.
We extend this construction to obtain quasi-cyclic codes.
\begin{defi}
\label{defi:vander}
Let $C_0,C_2,\ldots,C_{k-1}$ be $k$ linear codes of the same length $n$ over $\F_q$,
and $A$ be a $k\times n$ matrix with entries in $\F_q$.
The matrix product codes are then defined as
\[
[C_0, C_1, \ldots,C_{k-1}]A=\{(c_0,c_2,\ldots,c_{k-1})A, c_0\in C_0,\ldots,c_{k-1}\in C_{k-1}\}.
\]
\end{defi}
\begin{rem}
\label{rem:vander}
Assume that the finite field $\F_q$ contains a unit of order $2^a$.
From Definition \ref{defi:vander} and (\ref{eq:Vander}),
the Vandermonde product of $C_0, C_1, \ldots, C_{2^a-1}$ for $k=2^a$ and $n=2^am'$ is the matrix product code
\[
[C_0, C_1, \ldots,C_{2^a-1}]A=\{(c_1,c_2,\ldots,c_{2^a-1})A, c_1\in C_1,\ldots,c_{2^a-1}\in C_{2^a-1}\},
\]
where
\begin{equation}
\label{matrix:vander}
A=\left(
  \begin{array}{cccc}
    1 & 1 & \cdots & 1 \\
    1 & \alpha^{-1}& \cdots & \alpha^{-2^a} \\
    \vdots & \vdots & & \vdots \\
    1 & \alpha^{-(2^a-1)} & \cdots & \alpha^{-(2^a-1)^{2^a}} \\
  \end{array}
\right).
\end{equation}
\end{rem}

The main result of this section is as follows.
\begin{thm}
Let $C_0, C_1, \ldots, C_{2^a-1}$ be linear codes of length $l$ over a finite field $\F_q$ which contains a unit of order $2^a$.
Then the Vandermonde product of $C_0,\ldots,C_{2^a-1}$ is a matrix product code with matrix $A$ given in (\ref{matrix:vander}), and
is a  quasi-cyclic code of length $2^{2a}m'$ and index $2^a$ over $\F_q$.
Moreover, every $2^a$-QC code of length $2^{2a}m'$ over $\F_q$ is obtained via this matrix product construction.
\end{thm}
\pf
Assuming the hypothesis of the theorem, it was proven in \cite[Theorem 6.4]{sole1} that the Vandermonde product of $C_0,\ldots,C_{2^a-1}$
is a quasi-cyclic code of length $2^{2a}m'$ and index $2^a$ over $\F_q$.
Moreover, every $2^a$-QC code of length $2^{2a}m'$ over $\F_q$ can be obtained via the Vandermonde construction.
The result then follows from Remark \ref{rem:vander}.
\qed

\begin{ex}
If $\F_q$ is a finite field with odd characteristic, then $\alpha=-1$ is a $2$-nd root of unity and so
the Vandermonde product of codes $C_1$ and $C_2$ of length $l$ over $\F_q$ is given by
\[
C=C_1\vee C_2 =\{(u+v|u-v),\;\;u\in C_1,\;\;v\in C_2 \}.
\]
If $G_1$ and $G_2$ are generator matrices of $C_1$ and $C_2$, respectively, then
\[
\left(
\begin{array}{cc}
G_1 & G_1 \\
G_2 & -G_2\\
\end{array}
\right),
\]
is a generator matrix of $C$,
so $C$ is an $l$-QC code of length $2l$.
\end{ex}

In the following, the Vandermonde product and the matrix product are used to
construct isodual quasi-cyclic codes of length $2^{2a}m'$.
\begin{cor}
\label{thm:vandermonde}
Let $\F_q$ be a finite field, $m'$ an odd integer such that
$(m',q)=1$, $q \equiv 1 \bmod 2^a$,
and $C_0,\ldots,C_{2^a-1}$ be linear isodual codes of length $2^am'$ over $\F_q$.
Then the Vandermonde product of $C_0,\ldots,C_{2^a-1}$
is an isodual quasi-cyclic code of length $2^{2a}m'$ and index $2^a$ over $\F_q$.
Moreover, every $2^a$-QC code of length $2^{2a}m'$ which is isodual over $\F_q$
can be obtained via the Vandermonde construction.
\end{cor}
\pf
If $q \equiv 1 \bmod 2^a$, then by \cite[Lemma 3.1]{BGGI} there exists a $2^a$-th root of unity in $\F_q$ and
$\alpha$ such that $\alpha^{2^a}=1$, so
the polynomial $Y^{2^a}-1$ decomposes completely in $\F_q[Y]$ as $\prod_{i=0}^{2^a-1}(Y-\alpha^i)$.
Then $\F_q[y]/\langle Y^{2^a}-1\rangle$ is isomorphic to the direct sum
$\bigoplus_{i=0}^{2^a-1}\F_q[Y]/\langle Y-\alpha^i\rangle$,
and since $\F_q[Y]/\langle Y-\alpha^i\rangle\simeq \F_q$,
by Theorem \ref{Thm:isodual} any $2^a$-QC code of length $l$ over $\F_q$ is isodual
if and only if it is a direct sum of $2^a$ linear isodual codes of length $l$ over $\F_q$.
\qed

If $G_0,G_1,\ldots,G_{2^a-1}$ are generator matrices of $C_0,\ldots,C_{2^a-1}$, respectively,
then their Vandermonde product is generated by
\[
G=\left(
  \begin{array}{cccc}
    G_0 & G_0 & \cdots & G_0 \\
    G_1 & \alpha^{-1}G_1& \cdots & \alpha^{-2^a}G_1 \\
    \vdots & \vdots & & \vdots \\
    G_{2^a-1} & \alpha^{-(2^a-1)}G_{2^a-1} & \cdots & \alpha^{-(2^a-1)^{2^a}}G_{2^a-1} \\
  \end{array}.
\right)
\]
As an example of the Vandermonde construction, consider the case when $m=2$ and the $(u+v|u-v)$ construction.
From Example \ref{ex:Herm}, we have that if $q$ is odd then an $l$-QC code $C$ of length $2l$
can be expressed as $C_1 \oplus C_2$ where $C_1$ and $C_2$ are linear codes of length $l$ over $\F_q$.
Moreover, $C$ is an isodual code if and only if $C_1$ and $C_2$ are isodual codes.
Hence from Corollary \ref{thm:vandermonde} we obtain the following result.
\begin{cor} Let $\F_q$ be a finite field and $q$ a power of an odd prime.
If $C_1$ and $C_2$ are isodual cyclic codes of length $l$ over $\F_q$ then the code
\[
C=C_1\vee C_2 =\{(u+v|u-v),\;\;u\in C_1,\;\;v\in C_2 \},
\]
is an isodual $l$-QC code of length $2l$ over $\F_q$.
Furthermore, all isodual $l$-QC codes of length $2l$ over $\F_q$ can be constructed in this way.
\end{cor}

If $G_i$ is the generator matrix of $C_i $, $i=1,2$, then
\[
G= \left(
\begin{array}{cc}
G_1 & G_1 \\
G_2 & -G_2\\
\end{array}
\right),
\]
is a generator matrix of $C$.

In the following, we consider the constructions of isodual cyclic codes over finite fields first presented in \cite{BGGI}.
Let $q$ be a power of an odd prime $p$ such that $q \equiv 1 \bmod 2^a$,
$a\geq 1$ an integer, $m'$ an odd prime, and $\alpha\in\F_q^*$ a primitive $2^a$-th root of unity.
\begin{enumerate}
\item [1)] If $f(x)$ is a polynomial in $\F_q[x]$ such that
\[
x^{m'}-1=(x-1)f(x),
\]
then the cyclic codes of length $2^am'$ generated by
\[
(x^{2^{a-1}}-1)\prod_{k=0}^{2^{a-1}-1} f(\alpha^{-2k-1}x),
\]
and
\[
(x^{2^{a-1}}+1)\prod_{k=1}^{2^{a-1}} f(\alpha^{-2k}x),
\]
are isodual codes of length $2^am'$ over $\F_q$.
\item [2)] If $f_1(x)$ and $f_2(x)$ are polynomials
in $\F_q[x]$ such that
\[
x^{m'}-1=(x-1)f_1(x)f_2(x),
\]
then the cyclic codes of length $2^am'$ generated by
\[
(x^{2^{a-1}}-1)\prod_{k=1}^{2^{a-1}} f_i(\alpha^{-2k}x)\prod_{k=0}^{2^{a-1}-1}f_j(\alpha^{-2k-1}x),
\]
and
\[
(x^{2^{a-1}}+1)\prod_{k=1}^{2^{a-1}} f_i(\alpha^{-2k}x)\prod_{k=0}^{2^{a-1}-1}f_j(\alpha^{-2k-1}x),
\]
$i,j\in\{1,2\}, i\neq j$, are isodual codes of length $2^am'$ over $\F_q$.
\item [3)] If there exists a pair of odd-like duadic codes $ D_{i}= \langle f_{i}(x) \rangle $ of odd length $m'$,
then we have the following codes.
\begin{enumerate}
\item [i)] The cyclic codes $C_{ij}$ and $C_{ij}'$ of length $2^am'$ over $\F_q$ generated by
\[
(x^{2^{a-1}}-1)\prod_{k=1}^{2^{a-1}} f_i(\alpha^{-2k}x)\prod_{k=0}^{2^{a-1}-1}f_j(\alpha^{-2k-1}x),
\]
and
\[
(x^{2^{a-1}}+1)\prod_{k=1}^{2^{a-1}} f_i(\alpha^{-2k}x)\prod_{k=0}^{2^{a-1}-1}f_j(\alpha^{-2k-1}x),
\]
$i,j\in\{1,2\}, i\neq j$, are isodual codes of length $2^am'$ over $\F_q$.
\item [ii)] If the splitting modulo $m'$ is given by $\mu_{-1}$, then the cyclic codes of
length $2^am'$ generated by
\[
(x^{2^{a-1}}-1)\prod_{k=1}^{2^{a}} f_i(\alpha^{-k}x),
\]
and
\[
(x^{2^{a-1}}+1)\prod_{k=1}^{2^{a}} f_i(\alpha^{-k}x),
\]
are isodual codes of length $2^am'$ over $\F_q$.
\end{enumerate}
\end{enumerate}
Using these constructions, we give some examples of isodual quasi-cyclic codes obtained from isodual cyclic codes over finite fields.
\begin{ex}
Over $\F_3$, we have $x^7-1 =(x + 2)(x^6 + x^5 + x^4 + x^3 + x^2 + x + 1)$, so that
\[
x^{14}-1= (x +2)(x^6 + x^5 + x^4 + x^3 + x^2 + x + 1)(x + 1)(x^6 - x^5 + x^4 - x^3 + x^2 - x + 1).
\]
The cyclic codes generated by
\[
g_0(x)=(x +2)(x^6 - x^5 + x^4 - x^3 + x^2 - x + 1),
\]
and
\[
g_1(x)=(x + 1)(x^6 + x^5 + x^4 + x^3 + x^2 + x + 1),
\]
are isodual cyclic codes of length $14$ with minimum distance $4$.
If
$C_0=\langle g_0(x) \rangle$ and $C_1=\langle g_1(x) \rangle $ with generator matrices $G_0$ and $G_1$, respectively,
then $C_0\vee C_1$ is an isodual $14$-QC code of length $28$ over $\F_3$ with generator matrix
\[
G= \left(
\begin{array}{cc}
G_0 & G_0 \\
G_1 & -G_1\\
\end{array}\right).
\]
\end{ex}
\begin{ex}
Over $\F_5$, $\beta=2$ satisfies $\beta^4=1$.
If $m=11$, we have
\[
x^{11}-1 = (x-1)(x^5+2x^4+4x^3+x^2+x+4)(x^5+4x^4+4x^3+x^2+3x+4),
\]
so that
\[
x^{11}-1 = (x-1)f_1(x)f_2(x),
\]
and
\[
x^{44}-1 = (x-1)f_1(x)f_2(x)(x+1)f_1(-x)f_2(-x)(x+2)f_1(2x)f_2(2x)(x-2)f_1(-2x)f_2(-2x).
\]
The cyclic codes generated by
\[
\begin{array}{ccl}
g_0(x)&=&(x^2-1)f_1(x)f_1(-x)f_1(2x)f_1(-2x),\\
g_1(x)&=&(x^2+1)f_1(x)f_1(-x)f_1(2x)f_1(-2x),\\
g_2(x)&=&(x^2-1)f_2(x)f_2(-x)f_2(2x)f_2(-2x),
\end{array}
\]
and
\[
\begin{array}{ccl}
g_3(x)&=&(x^2+1)f_2(x)f_2(-x)f_2(2x)f_2(-2x),
\end{array}
\]
are isodual cyclic codes of length $44$ over $\F_5$.
If $C_i=\langle g_i(x)\rangle$ with generator matrix $G_i$, $i=0,1,2,3$,
then $C_0\vee C_1\vee C_2\vee C_3$ is a $44$-QC isodual code of length $176$ over $\F_5$
with generator matrix
\[G=\left(
      \begin{array}{cccc}
        G_0& G_0& G_0 & G_0 \\
        G_1 & -2G_1 & -1G_1 & 2G_1 \\
        G_2& -G_2 & G_2 & -G_2\\
        G_3& 2G_3 & -G_3 & -2G_3\\
      \end{array}
    \right).
\]
\end{ex}

\subsection{Cubic Isodual Codes}
In this section, it is assumed that $m=3$ and $q$ is not a power of $3$.
We know that if $q\equiv 2 \bmod 3$, then $Y^2+Y+1$ is irreducible in $\F_q[Y]$ so that
\[
Y^3-1=(Y-1)(Y^2+Y+1),
\]
is a product of irreducible factors.
By (\ref{eq:decom2}), we have the decomposition
\[
R=\frac{\F_{q}[Y]}{\langle Y^3-1\rangle}=\F_q\oplus \F_{q^2}.
\]
This gives a correspondence between $l$-QC codes $C$ of length $3l$ over $\F_q$
and a pair of codes $(C_1,C_2)$
where $C_1$ is a linear code over $\F_q$ of length $l$ and $C_2$ is a linear code
over $\F_{q^2}$ of length $l$.
By \cite[Theorem 5.1]{sole1}, we have
\begin{equation}
\label{eq:cubic}
C=\{(x+2a-b|x-a+2b|x-a-b)\,|x\in C_1,\,a+\beta b\in C_2\},
\end{equation}
where $\beta^2+\beta+1=0$.
Moreover, by Theorem \ref{Thm:isodual} $C$ is an isodual code over $\F_q$ if and only if $C_1$ is an
isodual code over $\F_q$ with respect to the Euclidean inner product and $C_2$ is an isodual code
over $\F_{q^2}$ with respect to the Euclidean inner product.

\begin{ex}
For $q = 5$ and $l=11$
\[
x^{11}-1 = (x-1)(x^5+2x^4+4x^3+x^2+x+4)(x^5+4x^4+4x^3+x^2+3x+4),
\]
so that
$x^{11}-1 = (x-1)f_1(x)f_2(x) = (x-1)f_1(x)f_1^*(x)$.
Then by \cite[Theorem 5.4]{BGGI}, the code $C_1$ of length $22$ over $\F_5$ generated by
$g_1(x)= (x-1)f_i(x)f_i^*(-x)$ is an isodual cyclic code with minimum distance $8$.
Further, the code $C_2$ of length $22$ over $\F_5$ generated by
$g_2(x)= (x+1)f_i(x)f_i(-x)$ is an isodual cyclic code with minimum distance $6$.
The same factorization of $x^{11}-1$ is obtained over $\F_{25}$,
so the codes $C'_1$ and $C'_2$ of length $22$ over $\F_{25}$ generated by $g_1(x)= (x-1)f_i(x)f_i^*(-x)$ and
$g_2(x)= (x+1)f_i(x)f_i(-x)$, respectively, are isodual cyclic codes.
Then the code
\[
C=\{(x+2a-b|x-a+2b|x-a-b)\,|x\in C_i,\,a+\beta b\in C'_i,\,1\leq i\leq2\},
\]
is an isodual QC code of length $3 \times 22$ over $\F_5$.
\end{ex}

\section{Conclusion}
In this paper, conditions on the equivalence of quasi-cyclic codes over finite fields were given.
Necessary and sufficient conditions for a quasi-cyclic code to be
isodual were presented using the properties of the constituent codes.
These conditions were used to obtain two constructions for isodual quasi-cyclic codes over finite fields
considering the Euclidean inner product of the isodual constituent codes.
Further, the matrix product was used to obtain isodual codes over finite fields.
Decoding up to half the minimum distance for these quasi-cyclic codes is possible
using a generalization of the decoding algorithm for matrix product codes \cite{hernando}.
Some of the results given in this paper can easily be generalized to quasi-cyclic codes over rings.
A more challenging problem is to obtain a general construction of quasi-cyclic codes as matrix product codes.


\end{document}